\documentclass[a4paper]{article}
\usepackage{graphicx}
\usepackage{amsmath}
\usepackage{amsfonts}
\usepackage{fancybox}
\usepackage{a4}

\def\Tr{\mathrm{Tr}}

\def\tr{\mathrm{tr}}

\def\half{{1\over2}}
\def\wt{\widetilde}

\def\[{\left[}
\def\]{\right]}
\def\({\left(}
\def\){\right)}

\def\cH{{\cal H}}

\def\cN{{\cal N}}
\def\cO{{\cal O}}

\def\bZ{{\mathbf Z}}

\def\mfi{\mathfrak{i}}

\newcommand{\ineq}{{}^{>}\!\!\!\> \!\!{}_{<}}

\def \be {\begin{equation}}
\def \ee {\end{equation}}
\def \bea {\begin{eqnarray}}
\def \eea {\end{eqnarray}}
\def \nn {\notag\\}

\usepackage[vcentermath]{youngtab}
\begin{document}

\begin{titlepage}
\title{
\begin{flushright}
\normalsize{ SNUTP11-009\\
TIFR/TH/11-43\\
Oct 2011}
\end{flushright}
       \vspace{2cm}
Index for Three Dimensional Superconformal Field Theories
and Its Applications
       \vspace{2cm}}
\author{
Shuichi Yokoyama\thanks{E-mail: \tt yokoyama[at]theory.tifr.res.in}
\\[30pt] 
{\it Department of Physics and Astronomy and Center for Theoretical Physics,}\\
{\it Seoul National University, Seoul 51-747, Korea}\\
{\it Department of Theoretical Physics, Tata Institute of Fundamental Research,}\\
{\it  Homi Bhabha Rd, Mumbai 400005, India }
}
\date{}

\maketitle
\thispagestyle{empty}

\vspace{0cm}

\begin{abstract}
\normalsize
We review aspects of superconformal indices in three dimension. 
Three dimensional superconformal indices 
can be exactly computed by using localization method 
including monopole contribution, and
can be applied to 
provide evidence for mirror duality, AdS$_4$/CFT$_3$ correspondence 
and global symmetry enhancement of strongly coupled gauge theories. 
After reviewing, 
we discuss the possibility of global symmetry enhancement 
in a finite rank of gauge group.
\end{abstract}

\end{titlepage}

\section{Introduction}

Duality is a powerful tool to study a strongly coupled gauge theory. 
Duality enables us to map strongly coupled region in a gauge theory to 
weakly coupled region in the dual theory, 
so phenomena such as color confinement can be addressed by perturbation from the dual theory. 
Several examples are known such as
S duality, Seiberg duality \cite{Seiberg:1994pq}, mirror duality \cite{Intriligator:1996ex}, 
AdS/CFT duality \cite{Maldacena:1997re}.

Although it is often too difficult to prove duality completely, 
one can provide evidence for duality by using a superconformal index. 
A superconformal index can be a test of duality 
by checking whether indices of both theories agree or not. 
Such an analysis was first carried out for the simplest example of AdS$_5$/CFT$_4$ duality, 
$\cN=4$ super Yang-Mills theory and its gravity dual \cite{Kinney:2005ej}.
The agreement of indices of both theories was confirmed in the large $N$ limit.
Four dimensional indices can be calculated for less supersymmetric gauge theories 
and they were applied to various types of duality
\cite{Romelsberger:2005eg,Nakayama:2005mf,Nakayama:2006ur,Gadde:2010en,Benvenuti:2006qr,Romelsberger:2007ec,Dolan:2008qi,Spiridonov:2008zr,Spiridonov:2011hf}.

Similarly, 
three dimensional indices can be computed. 
Index calculation in three dimension has been performed actively 
since $\cN=6$ Chern-Simons theory (ABJM theory) appeared \cite{Aharony:2008ug},
which provides the first example of AdS$_4$/CFT$_3$ duality. 
An index for ABJM theory was first computed by taking 't Hooft limit \cite{Bhattacharya:2008bja},
which excludes monopole contribution. 
It was confirmed that the index agrees with that of dual type IIA theory. 
The index for ABJM theory including monopole contribution can be also calculated \cite{Kim:2009wb}.
It was checked that it coincides with that of dual M-theory in the large $N$ limit. 
This kind of analysis can be done for ${\cal N}=5,4,3$ Chern-Simons theories 
\cite{Choi:2008za,Imamura:2009hc,Kim:2010vwa}. 
An index for a general ${\cal N}=2$ superconformal field theories was first computed 
in \cite{Imamura:2011su}, 
and it can be applied to AdS$_4$/CFT$_3$ duality \cite{Imamura:2011uj,Cheon:2011th},
mirror duality \cite{Imamura:2011su} with refined analysis \cite{Krattenthaler:2011da,Kapustin:2011jm}
and other types of duality 
\cite{Jafferis:2011ns,Bashkirov:2011vy,Hwang:2011qt,Gang:2011xp,Hwang:2011ht}.

In the next section, 
we make a brief review of a derivation of a three dimensional superconformal index
and give an index formula explicitly. 
From this formula, 
we can obtain a formula of an index in the large $N$ limit. 
In Section \ref{apply}, 
we apply the formula to several $\cN=2$ gauge theories. 
We will give evidence for mirror duality and AdS$_4$/CFT$_3$ duality 
by using indices. 
The final section is devoted to summary and discussion.

\section{Superconformal index}

The definition of a superconformal index is 
\bea
I&=& \Tr_{\cH} \biggl((-1)^F e^{-\beta'\{Q, S\}} \, C \biggr). 
\label{ind}
\eea
Here $\cH$ is the Hilbert space of the theory, $F$ is fermion number operator, 
$Q$ is a supercharge satisfying the nilpotent condition, 
$S$ is the hermitian conjugate of $Q$, 
$C$ is a chemical potential which commutes with $Q$, 
and $\beta'$ is a parameter.  
Indices are defined so that 
they receive contribution only from supersymmetric states. 
This is due to the fact that 
in a supersymmetric theory 
there is one-to-one correspondence between bosons and fermions 
in non-supersymmetric states 
and their contributions totally cancel thanks to insertion of $(-1)^F$ \cite{Witten:1982df}. 
As a result, \eqref{ind} is independent from $\beta'$.
In this section, 
we will show a general formula of a three-dimensional superconformal index 
and the large $N$ version of the formula. 

\subsection{Formula in three dimension} 

Before showing a formula of an index, 
we make a brief review on how to compute an index.
To calculate an index, 
we have to specify the Hilbert space of the theory. 
For this purpose, 
since the theory is a euclidean conformal field theory, 
we use the radial quantization.
In other words, 
we consider the theory not on $R^3$ but on $S^2 \times R$ 
by using a conformal transformation. 
Here $R$ is the radial direction of $R^3$. 
By regarding this radial direction as time, 
we quantize the theory and obtain the Hilbert space as a set of the quantized states.

We compute a superconformal index of this theory in path-integral approach 
by using a localization method \cite{Kim:2009wb,Imamura:2011su}. 
To do this, 
we first compactify the radial direction $R$ to $S^1$ to read off the charge assignment of the states. 
Then we add a $Q$-exact term into the original Lagrangian. 
Since only BPS states contribute to the index, 
this $Q$-exact deformation does not change the index at all. 
The point is that 
we can choose a $Q$-exact term in such a way that 
it becomes kinetic terms of the fields on $S^2\times S^1$ \cite{Kim:2009wb,Kapustin:2009kz}. 
Then we look for saddle points of the $Q$-deformed action. 
It turns out that
saddle points are given by GNO monopole solutions \cite{Goddard:1976qe}, 
which is specified by a magnetic flux through $S^2$, 
with holonomy, which is a zero-mode of a gauge field on $S^1$.
Note that saddle points are sensitive to the topology of the space. 
Then we expand the theory around a saddle point 
and take the weak coupling limit, 
which can be realized by sending the coupling of the $Q$-exact term to infinity. 
After the weak coupling limit, 
the total Lagrangian consists of quadratic terms of the fields on a monopole background 
and thus we can perform path-integral exactly, which reduces to gaussian integrals. 
The index is decomposed in the following form. 
\begin{equation}
I=\sum_M t^M\int\! d\alpha e^{-S_{0}} \biggl[\prod_{v} I^{v} \prod_{q} I^q\biggr]. 
\label{indR3}
\end{equation}
Here we introduce the chemical potential $t$ for a magnetic charge $M$ in a collective notation. 
We sum the contribution over all saddle points parametrized by a magnetic charge $M$ and 
holonomy $\alpha$. 
$e^{-S_{0}}$ is the contribution coming from the original action, 
which does not vanish only when it includes a Chern-Simons term
\begin{equation}
S_{0}
= \sum_v i k_v \tr_v( \alpha M), 
\end{equation}
where $k_v$ is the Chern-Simons coupling of a vector multiplet $v$ 
and $\tr_v$ is trace for the gauge group associated with $v$.
$I^v$ and $I^q$ are the contributions coming from a vector multiplet and a chiral multiplet, 
respectively.

The result of the path-integral is the following.
The contribution of a vector multiplet is \cite{Kim:2009wb}
\be
I^v = 
{1 \over \mbox{(sym)}}
\prod_{\substack{\hat v(M)=0}} 2 i\sin\left({\hat v(\alpha)\over 2}\right)\prod_{\hat v} \biggl[
x^{-\half |\hat v(M)|}
\exp\left[
\sum_{m=1}^\infty\frac{1}{m} \mfi^v(e^{im \alpha},x^m)
\right]\biggr].
\ee
$\hat v$ represents a weight of the gauge representation of the vector multiplet $v$.
The factors ${1 \over \mbox{(sym)}}$
and 
$\prod_{\substack{\hat v(M)=0}} 2 i\sin\left({\hat v(\alpha)\over 2}\right)$ 
come from the Weyl symmetry and gauge fixing of the unbroken gauge group, respectively. 
When the gauge group $G$ is broken to $\prod_i G_i$ by monopole flux,
(sym) is given by $\prod_i (\mbox{rank} G_i)!$.  
The factor $x^{-\half |\hat v(M)|}$ describes the zero-point energy of a vector multiplet on a monopole background.
$\mfi^v$ describes a single excitation of a vector multiplet on a monopole background, called a letter index. 
\begin{equation}
\mfi^v(x,e^{i\alpha})=
-e^{i\hat v(\alpha)}x^{|\hat v(M)|}. 
\label{fvector}
\end{equation}
The following exponential 
\begin{equation}
\exp\left[\sum_{m=1}^\infty\frac{1}{n} {\mathfrak i}(e^{im\alpha},x^m,z_i^m)\right]
\label{excitation}
\end{equation}
is called plethystic exponential. 
The plethystic exponential of a letter index describes an index of a multi-excitation.

The contribution of a chiral multiplet is  \cite{Imamura:2011su}
\be
I^q = \prod_{\hat q} \biggl[
e^{-\frac{i}{2} |\hat q (M)|\hat q (\alpha)}x^{\half(1-\Delta_q)|\hat q (M)|}z_i^{-\half|\hat q (M)|F_i} 
\exp\left[
\sum_{m=1}^\infty\frac{1}{m} \mfi^q(e^{im \alpha},x^m,z_i^n)
\right],
\ee
where $\hat q $ represents a weight of the gauge representation of the chiral superfield $q$. 
The structure is the same as that of a vector multiplet. 
The first factor in front of plethystic exponential describes the contribution of 
zero-point fluctuation of a chiral multiplet. 
$\mfi^q$ is a letter index of a chiral multiplet.
\begin{equation}
\mfi^{q}(e^{i\alpha},x,z_i)=
e^{i\hat q (\alpha)}z_i^{F_i}\frac{x^{|\hat q (M)|+\Delta_q}}{1-x^2}
-e^{-i\hat q (\alpha)}z_i^{-F_i}
\frac{x^{|\hat q (M)|+2-\Delta_q}}{1-x^2}, 
\label{fchiral}
\end{equation}
where $\Delta_q$ is a conformal dimension of the scalar field in a chiral multiplet.
Plethystic exponential describes a multi-excitation of a chiral multiplet on a monopole background.

When the conformal dimension is canonical, $\Delta_q=\half$, 
the formula of indices was derived in \cite{Kim:2009wb}. 
We generalized the result of \cite{Kim:2009wb} so that
it can be applicable to theories with non-canonical R-charge assignments \cite{Imamura:2011su}. 
It was pointed out that 
this formula can be further generalized to 
$\cN=2$ theories in a non-trivial background gauge field 
coupling to global symmetry currents \cite{Kapustin:2011jm}.

\subsection{Large $N$ formula}

From a general formula for an index shown in the previous section, 
we can obtain an index formula in the large $N$ limit \cite{Imamura:2011uj}. 
In application of the formula, 
gauge theories we have in mind are quiver gauge theories, 
so
we specify the gauge group as a product of unitary groups $\prod_v U(N_v)_v$ and 
representations of the chiral matters as bi-fundamental ones. 
It is not difficult to do for other types of gauge group and representation such as 
fundamental/anti-fundamental representations.
In this situation, the saddle points are labeled by 
the diagonal components of magnetic charges
$M^v_s$ and holonomy $\alpha^v_s$ for each gauge group $U(N_v)_v$,  
where $s=1, \cdots, N_v$.
So the formula is rewritten as
\begin{equation}
I(x,z_i)
=\sum_M t^M{1\over (\mbox{\rm sym})}  \int d\alpha e^{-S_{0}}
e^{i b_0(\alpha)}x^{\epsilon_0}z_i^{f_{0i}} 
\exp\left[\sum_{m=1}^\infty\frac{1}{n}\mfi (e^{ima},x^m,z_i^m)\right].
\label{formula}
\end{equation}
Here the total letter index $\mfi$ consists from those of vector and hyper multiplets, 
$\mfi=\sum_v  \mfi^v+\sum_q \mfi^q$, where
\begin{eqnarray}
&&\mfi^v(e^{i\alpha},x)
=
\sum_{s,t=1}^{N_v}
(1-\delta_{s,t})
\left(-e^{i(\alpha^{v}_s- \alpha^v_t)}x^{| M^v_s- M^v_t|}\right),
\label{fvecdef}\\
&&
\!\!\!\!\!\!\!\!\!\!\!\!\!\mfi^q(e^{i\alpha},x,z_i)
=
\sum_{s=1}^{N_{h_q}}
\sum_{t=1}^{N_{t_q}}
\frac{x^{|M^{h_q}_s-M^{t_q}_t|}}{1-x^2}
\left(
e^{i(\alpha^{h_q}_s-\alpha^{t_q}_t)}z_i^{F_i(q)}x^{\Delta(q)}
-e^{-i(\alpha^{h_q}_s-\alpha^{t_q}_t)}z_i^{-F_i(q)}x^{2-\Delta(q)}
\right),
\label{fchdef}
\end{eqnarray}
where we assume that the chiral multiplet $q$ is $(N_{h_q}, \bar N_{t_q})$ representation
for the gauge group $U(N_{h_q})_{h_q} \times U(N_{t_q})_{t_q}$.
The zero-point contributions is
\begin{eqnarray}
\epsilon_0
&=&\frac{1}{2}\sum_{q}
\sum_{s=1}^{N_{h_q}}
\sum_{t=1}^{N_{t_q}}
|M^{h_q}_s-M^{t_q}_t|(1-\Delta(q))
-\frac{1}{2}\sum_{v}
\sum_{s=1}^{N_v}\sum_{t=1}^{N_v}|M^{v}_s-M^{v}_t|,
\label{ep0}\\
f_{0i}
&=&-\frac{1}{2}\sum_{q}
\sum_{s=1}^{N_{h_q}}
\sum_{t=1}^{N_{t_q}}|M^{h_q}_s-M^{t_q}_t|F_i(q),
\label{q0i}
\\
b_0(\alpha)
&=&-\frac{1}{2}\sum_{q}
\sum_{s=1}^{N_{h_q}}
\sum_{t=1}^{N_{t_q}}
|M^{h_q}_s-M^{t_q}_t|
(\alpha^{h_q}_s-\alpha^{t_q}_t).
\label{b0}
\end{eqnarray}

The striking feature of the large $N$ index is that 
the index can be factorized into three parts by monopole charges as
\be
I= I^{(0)} I^{(+)} I^{(-)}.
\label{factor}
\ee
Here $I^{(0)}, I^{(+)}, I^{(-)}$ are the neutral, positive, negative parts of the index. 
Let us explain this below. 
To see this factorization, 
we can divide the letter index into three parts  
\be
\mfi = \mfi^{(0)}+\mfi^{(+)}+\mfi^{(-)}, 
\ee
where $\mfi^{(*)}= \sum_v \mfi^{v(*)}+\sum_q \mfi^{q(*)}$ and  
\bea
\!\!\!\!\!\!\!\!&&\!\!\!\!\!\!\!\!\!\!\!\!\!\mfi^{v(0)}=
\sum_{s,t=1}^{N_v}
\left(-e^{i(\alpha^{v}_s- \alpha^v_t)}x^{|M^v_s|+|M^v_t|}\right), \\
\!\!\!\!\!\!\!\!&&\!\!\!\!\!\!\!\!\!\!\!\!\!\mfi^{v(\pm)}=
\sum_{s,t=1}^{N^{(\pm)}_v}
e^{i(\alpha^{v}_s- \alpha^v_t)}\left(-(1-\delta_{s,t})x^{|M^v_s - M^v_t|}+ x^{|M^v_s|+|M^v_t|}\right), \\
\!\!\!\!\!\!\!\!&&\!\!\!\!\!\!\!\!\!\!\!\!\!\mfi^{q(0)}=
\sum_{s=1}^{N_{h_q}}
\sum_{t=1}^{N_{t_q}}
\frac{x^{|M^{h_q}_s|+|M^{t_q}_t|}}{1-x^2}
\left(
e^{i(\alpha^{h_q}_s-\alpha^{t_q}_t)}z_i^{F_i(q)}x^{\Delta(q)}
-e^{-i(\alpha^{h_q}_s-\alpha^{t_q}_t)}z_i^{-F_i(q)}x^{2-\Delta(q)}
\right),\\
\!\!\!\!\!\!\!\!&&\!\!\!\!\!\!\!\!\!\!\!\!\!\mfi^{q(\pm)}=
\sum_{s=1}^{N^{(\pm)}_{h_q}}
\sum_{t=1}^{N^{(\pm)}_{t_q}}
{ x^{|M^{h_q}_s-M^{t_q}_t|}- x^{|M^{h_q}_s|+|M^{t_q}_t|} \over 1-x^2}
\left(
e^{i(\alpha^{h_q}_s-\alpha^{t_q}_t)}z_i^{F_i(q)}x^{\Delta(q)}
-e^{-i(\alpha^{h_q}_s-\alpha^{t_q}_t)}z_i^{-F_i(q)}x^{2-\Delta(q)}
\right).
\eea
$\sum_{s=1}^{N^{(\pm)}_v}$ means summation over $s$ satisfying 
$M_s \ineq 0$, where $M>0 \;(M<0)$ means 
that $M$ is non-zero and the components are all non-negative (non-positive). 
This is due to the fact that 
$\mfi^{v(\pm)}$ and $\mfi^{q(\pm)}$ vanish unless 
$M^v_s, M^v_t$ and $M^{h_q}_s, M^{t_q}_t$ are the same signature, respectively. 

We can also divide the factors in front of plethystic exponential by three parts 
as follows. 
\bea
S_{0}&=&S_{0}^{(0)}+S_{0}^{(+)}+S_{0}^{(-)},\\
\epsilon_{0}&=&\epsilon_0^{(0)}+\epsilon_{0}^{(+)}+\epsilon_0^{(-)},\\
f_{0i}&=&f^{(0)}_{0i}+f_{0i}^{(+)}+f_{0i}^{(-)},\\
b_{0}&=&b^{(0)}_0+b_{0}^{(+)}+b_{0}^{(-)}.
\eea
The neutral sector is 
\bea
S_{0}^{(0)}&=&0,\\
\epsilon_0^{(0)}
&=&N \sum_v 
\left(\sum_{ h_q= v \atop t_q=v}{1-\Delta(q) \over 2} - 1\right) \sum_{s=1}^{N_v} |M^v_s|,
\label{ep00}\\
f_{0i}^{(0)}
&=&-N \sum_v 
\left(\sum_{h_q= v \atop t_q=v}{F_i(q) \over 2}\right) \sum_{s=1}^{N_v} |M^v_s|,
\label{fp00}\\
b_0^{(0)}(\alpha)
&=&-\frac{1}{2}\sum_{q}
\sum_{s=1}^{N_{h_q}}
\sum_{t=1}^{N_{t_q}}
\left(|M^{h_q}_s|+|M^{t_q}_t|\right)
(\alpha^{h_q}_s-\alpha^{t_q}_t), 
\label{b00}
\eea
The charged sector is 
\begin{eqnarray}
\!\!\!\!\!\!\!\!\!\!\!\!S_{0}^{(\pm)}&=&\sum_v\sum_{s=1}^{N^{(\pm)}_v} i k_v \alpha^v_s M^v_s,\\
\!\!\!\!\!\!\!\!\!\!\!\!\epsilon_0^{(\pm)}
&=&\frac{1}{2}\biggl(\sum_{q}
\sum_{s=1}^{N^{(\pm)}_{h_q}}
\sum_{t=1}^{N^{(\pm)}_{t_q}}
{\bf M}(h_q,t_q:s,t)(1-\Delta(q)) -
\sum_{v} \sum_{s=1}^{N^{(\pm)}_v}\sum_{t=1}^{N^{(\pm)}_v}
{\bf M}(v,v:s,t)\biggr),
\label{ep02}\\
\!\!\!\!\!\!\!\!\!\!\!\!f_{0i}^{(\pm)}
&=&-\frac{1}{2}\sum_{q}
\sum_{s=1}^{N^{(\pm)}_{h_q}}
\sum_{t=1}^{N^{(\pm)}_{t_q}}
{\bf M}(h_q,t_q:s,t) F_i(q),
\label{q0i2}\\
\!\!\!\!\!\!\!\!\!\!\!\!b_0(\alpha)
&=&-\frac{1}{2}\sum_{q}
\sum_{s=1}^{N^{(\pm)}_{h_q}}
\sum_{t=1}^{N^{(\pm)}_{t_q}}
{\bf M}(h_q,t_q:s,t)
(\alpha^{h_q}_s-\alpha^{t_q}_t).
\end{eqnarray}
Here we set 
\bea
{\bf M}(a,b:s,t)=|M^{a}_s-M^{b}_t| -|M^{a}_s|-|M^{b}_t|.
\eea
By using these decompositions, $I^{(0)}, I^{(\pm)}$ in \eqref{factor} is written as 
\bea
\!\!\!\!\!\!\!\!\!\!\!\!
I^{(0)}&=& \int d\alpha 
e^{i b^{(0)}_0}x^{\epsilon^{(0)}_0}z_i^{f^{(0)}_{0i}} 
\exp\left[\sum_{m=1}^\infty\frac{1}{n}\mfi^{(0)} (e^{ima},x^m,z_i^m)\right],\\
\!\!\!\!\!\!\!\!\!\!\!\!
I^{(\pm)}
&=&\sum_{M\ineq 0} t^M{1\over (\mbox{\rm sym})}  \int d\alpha e^{-S^{(\pm)}_{(0)}}
e^{i b^{(\pm)}_0}x^{\epsilon^{(\pm)}_0}z_i^{f^{(\pm)}_{0i}} 
\exp\left[\sum_{m=1}^\infty\frac{1}{n}\mfi^{(\pm)} (e^{ima},x^m,z_i^m)\right].
\eea

Let us assume $b_0(\alpha)$ vanishes as is the case with vector-like theories. 
Under this assumption, 
we can perform the holonomy integral in $I^{(0)}$ in the large $N$ limit.  
By using the following notation 
\begin{equation}
\lambda_{v,m}
=
\sum_{s=1}^{N_v}
(x^{| M_s|}
e^{i \alpha_s})^m,
\label{lambdadef}
\end{equation}
we can rewrite $\mfi^{(0)}$ as
\be
\mfi^{(0)} 
=-\sum_{v}\lambda_{A,+1}\lambda_{A,-1}
+\sum_{q}
\left[
\lambda_{h_q,+1}
\lambda_{t_q,-1}
\frac{z_i^{F_i}x^{\Delta(\Phi)}}{1-x^2}
-
\lambda_{h_q,-1}
\lambda_{t_q,+1}
\frac{z_i^{-F_i}x^{2-\Delta(\Phi)}}{1-x^2}
\right], 
\ee
which is quadratic of $\lambda_v$. 
Therefore, if we set 
$\mfi^{(0)} =-\sum_{v,v'}\lambda_{v,+1}M_{v,v'}(x,z_i)\lambda_{v',-1}$, 
then $I^{(0)}$ can be calculated as
\begin{eqnarray}
I^{(0)}
&=&\int d\lambda \; x^{\epsilon^{(0)}_0}z_i^{f^{(0)}_{0i}} 
\exp\left(-\sum_{m=1}^\infty\frac{1}{m}\sum_{v,v'}M_{v,v'}(x^m,z_i^m)
\lambda_{v,m}\lambda_{v',-m}
\right)
\nonumber\\
&=&x^{\epsilon^{(0)}_0}z_i^{f^{(0)}_{0i}}\prod_{m=1}^\infty\left(\det _{v,v'} M^{-1}_{v,v'}(x^m,z_i^m)\right).
\label{neutralint2}
\end{eqnarray}

\section{Applications}
\label{apply}
In the previous section, 
we showed a formula of a superconformal index with a general R-charge assignment. 
In this section, testing whether the formula works correctly, 
we apply it to ${\cal N}=2$ superconformal field theories 
which have a non-canonical R-charge assignment 
and show evidence of mirror duality and AdS$_4$/CFT$_3$ duality.

\subsection{Mirror duality}

We apply the formula to a mirror pair of ${\cal N}=2$ gauge theories. 
In this paper, we use a mirror pair studied in 
\cite{Aharony:1997bx,deBoer:1997ka}.

One of the pair we study is $U(1)$ supersymmetric Maxwell theory (SQED) with a fundamental flavor, 
which means a pair of fundamental/anti-fundamental chiral multiplet. 
This theory has flavor symmetry $U(1)_L\times U(1)_R$
and toplogical symmetry $U(1)_J$, which is generated by topological current 
$J^\mu= \varepsilon^{\mu\nu\rho} F_{\nu\rho}$.
It is easily seen that 
the diagonal $U(1)$ group of the flavor symmetry is included 
in the gauge symmetry,  
so it is sufficient to take account of $U(1)_{L-R}$, 
which act on two chiral fields in the same way.  
This theory flows to non-trivial fixed point in the infra-red (IR) region. 
Let us denote the $U(1)_R$ charge of the chiral fields as $h$.  
Then the letter index of the IR theory is 
\begin{equation}
\mfi_{\rm QED}(x,e^{i\alpha},y)=f_{h,M}(x,e^{i\alpha}y)+f_{h,M}(x,e^{-i\alpha}y), 
\end{equation}
where $y$ describes the chemical potential for $U(1)_{L-R}$. 
$f$ is defined by
\begin{equation}
f_{\Delta,M}(x,y)=\frac{yx^{|M|+\Delta}-y^{-1}x^{|M|+2-\Delta}}{1-x^2}.
\end{equation}
Notice that the contribution of a vector multiplet becomes trivial since the theory is abelian. 
The total index is given by 
\begin{equation}
I_{\rm QED}(x,y,t)=\sum_{M\in{\bf Z}}t^{M}\int\frac{d\alpha}{2\pi}x^{(1-h)|M|}y^{-|M|}
\exp\left(\sum_{n=1}^\infty\frac{1}{n}\mfi_{\rm QED}(x^n,e^{in\alpha},y^n)\right).
\end{equation}

The mirror dual theory is known as 
a Wess-Zumino model of three chiral multiplets $q, \wt q, S$ with cubic superpotential, $\wt q S q$. 
From this superpotential, 
we can see two global $U(1)$ symmetries assigned for three fields $q, \wt q, S$
as $1, -1, 0$ and $-1, -1, 2$, respectively. 
Due to permutation symmetry of three chiral fields, 
we can determine the conformal dimension of the fields as two thirds 
so that the superpotential has R-charge two. 
It will turn out that, however, 
this information is not necessary for indices of both theories to match. 
To see the agreement of both indices, 
we need the relation of conformal dimensions of chiral fields coming from 
the correspondence of chiral operators such that 
$\wt Q Q \leftrightarrow S$, where $Q, \wt Q$ are chiral fields of SQED. 
From this relation, 
the conformal dimensions of three chiral fields $q, \wt q, S$ are fixed as $1-h, 1-h, 2 h$, 
respectively. 
By combining these facts,  
we can obtain the letter index of the dual theory 
\begin{equation}
\mfi_{{\rm WZ}}(x,y',t')
=f_{1-h,0}(x,t'y'^{-1})
+f_{1-h,0}(x,t'^{-1}y'^{-1})
+f_{2h,0}(x,y'^2)
\end{equation}
and the total index 
\begin{equation}
I_{{\rm WZ}}(x,y',t')=
\exp\left(\sum_{n=1}^\infty\frac{1}{n}\mfi_{{\rm WZ}}(x^n,y'^n,t'^n)\right). 
\end{equation}
Here $t', y'$ are the chemical potentials for the two global $U(1)$ symmetries.

By using numerical calculation, 
we can show that both indices agree in a series expansion form 
by setting $t'=t, y'=y$ \cite{Imamura:2011su}.
\begin{align}
&
I_{\rm QED}(x,y,t)=I_{{\rm WZ}}(x,y,t)\nonumber\\
&=\left(1
+\left(\frac{1}{ty}+\frac{t}{y}\right)a
+\left(\frac{1}{t^2y^2}+\frac{t^2}{y^2}\right)a^2
+\left(\frac{1}{t^3y^3}+\frac{t^3}{y^3}\right)a^3
+\left(\frac{1}{t^4y^4}+\frac{t^4}{y^4}\right)a^4
+{\cal O}(a^5)\right)\nonumber\\
&+\left(y^2-2a^2+\frac{a^4}{y^2}+{\cal O}(a^5)\right)b^2+\left(y^4+\left(\frac{y}{t}+ty\right)a^3-3a^4+{\cal O}(a^5)\right)b^4+{\cal O}(b^6), 
\end{align}
where we set 
$a=x^{1-h}, b=x^h.$

One can show the agreement of both indices analytically \cite{Krattenthaler:2011da}. 
Let us explain the analytic proof briefly. 
For this purpose, 
it is convenient to define q-shifted factorials following \cite{Krattenthaler:2011da}.
\be
(A;q)_{n} = \biggl\{
\begin{array}{lc}
\prod_{k=0}^{n-1} (1-A q^k)  & (n>0)\\
1  & (n=0)\\
\prod_{k=0}^{|n|-1} (1-Aq^{-k-1})^{-1}  & (n<0).
\end{array}
\ee
We also use the abbreviation such that $(A;q)=(A;q)_{\infty}$ and $(A_1, A_2, \cdots, A_l;q)
=(A_1;q) (A_2;q)  \cdots (A_l;q)$. 
By using this notation, the indices of SQED and the dual Wess-Zumino model can be rewritten as
\bea
\!\!\!\!\!\!\!\!\!\!\!\!\!\!\!\!I_{\rm QED}(x,y,t)
&=&\sum_{M\in{\bf Z}}t^{M}\oint\frac{dz}{2\pi i z} a^{|M|/2}y^{-|M|}
{(z^{-1}y^{-1}a^\half x^{|M|+1}, zy^{-1}a^\half x^{|M|+1};x^2)
\over 
(zya^{-\half} x^{|M|+1}, z^{-1}ya^{-\half} x^{|M|+1};x^2)},
\label{sqed}\\
\!\!\!\!\!\!\!\!\!\!\!\!\!\!\!\!I_{\rm WZ}(x,y',t')
&=&{(t'^{-1}y' a^{-\half} x^{2}, t' y' a^{-\half} x^{2}, y'^{-2} a ;x^2)
\over 
(t'y'^{-1} a^{\half} , t'^{-1} y'^{-1} a^{\half}, y'^2 a^{-1} x^2 ;x^2)}.
\label{wz}
\eea
The contour integral in \eqref{sqed} is carried out around the unit circle. 
The poles of the integrand in \eqref{sqed} 
in the unit circle are $z= y a^{-\half} x^{|M|+1+j}$, where $j\in\bZ\geq0$ 
when $x<a^{\half}/z<1$. 
Picking up these poles, 
one can perform the residue calculus of \eqref{sqed} 
\be
I_{\rm QED}(x,y,t)
=\sum_{M\in{\bf Z}}t^{M} \hat a^{|M|/2}
\sum_{j=0}^\infty {(\hat a x^{-2j}, x^{2(|M|+1+j)};x^2)
\over 
(\hat a^{-1} x^{2(|M|+1+j)}, x^{2};x^2) (x^{-2j};x^2)_j},
\ee
where we set $\hat a = y^{-2}a$. 
It turns out from a straight forward calculation 
that one can perform the summation over $M$ without taking the absolute value for $M$. 
By using this fact and a little calculation, 
one finds 
\be
I_{\rm QED}(x,y,t)
=
\sum_{j=0}^\infty \sum_{M\in{\bf Z}}(t\hat a^{\half})^M {(\hat a^{-1} x^{2(1+j)};x^2)_M
\over 
(x^{2(1+j)};x^2)_M}
{(\hat a x^{-2j}, x^{2(1+j)};x^2)
\over 
(\hat a^{-1} x^{2(1+j)}, x^{2};x^2) (x^{-2j};x^2)_j}. 
\ee
By using Ramanjujan summation formula
\be
\sum_{M\in\bZ} {(A;q)_M \over (B;q)_M} z^M = 
{(q, B/A, Az, q/(Az);q) \over (B,q/A,z, B/(Az);q)},
\ee
one can perform the summation over $M$ as 
\bea
I_{\rm QED}(x,y,t)
&=&
{(\hat a ;x^2)\over (t \hat a^{\half}, t^{-1} \hat a^{\half};x^2)}
\sum_{j=0}^\infty {( t^{-1} \hat a^{\half} x^{-2j}, t \hat a^{-\half} x^{2(1+j)};x^2)
\over 
(\hat a^{-1} x^{2(1+j)};x^2) (x^{-2j};x^2)_j}\\
&=&
{(\hat a, t \hat a^{-\half}x^2 ;x^2)\over (t \hat a^{\half}, \hat a^{-1}x^2  ;x^2)}
\sum_{j=0}^\infty {(  \hat a^{-1} x^2 ;x^2)_j \over (x^{2};x^2)_j} (t^{-1} \hat a^{\half})^j.
\eea
By using binomial theorem
\be
\sum_{j=0}^\infty {(A;q)_j \over (q;q)_j} z^j = 
{(Az;q) \over (z;q)},
\ee
one can also carry out the summation over $j$ as 
\bea
I_{\rm QED}(x,y,t)
&=&
{(\hat a, t \hat a^{-\half}x^2, t^{-1} \hat a^{-\half}x^2 ;x^2)\over (t \hat a^{\half},  \hat a^{-1}x^2, t^{-1}\hat a^{\half};x^2)},
\eea
which precisely agrees with $I_{\rm WZ}(x,y,t)$ in \eqref{wz}.

One can study mirror pairs with more flavor case \cite{Imamura:2011su,Krattenthaler:2011da} 
and with a more general background \cite{Kapustin:2011jm}.

\subsection{AdS$_4$/CFT$_3$ duality}
\label{ads/cft}

We apply the formula to an $\cN=2$ Chern-Simons-matter theory 
which has M-theory dual on AdS$_4 \times Q^{1,1,1}$, 
where $Q^{1,1,1}$ is a homogeneous manifold defined by
\begin{equation}
\frac{SU(2)\times SU(2)\times SU(2)}{U(1)\times U(1)}. 
\end{equation}
From the definition, 
$Q^{1,1,1}$ has the isometry
\begin{equation}
SU(2)_1\times SU(2)_2\times SU(2)_3\times U(1)_R.
\label{isom}
\end{equation}
The last $U(1)_R$ factor is identified with the R-symmetry.

A corresponding gauge theory 
is proposed as a quiver Chern-Simons theories studied in \cite{Franco:2008um}. 
The field contents and symmetries are 
shown in Table \ref{table:q111}.
\begin{table}[htb]
\caption{Symmetries and their charge assignments for the fields of a gauge theory dual to M-theory dual on AdS$_4 \times Q^{1,1,1}$.
In the table, $U(N)_{a,k_a}\,(a=1,2,3,4)$ is $a$-th $U(N)$ gauge group with Chern-Simons level $k_a$, 
$U(1)_i\,(i=1,2,3)$ is Cartan subgroup of $SU(2)_i$. }
\label{table:q111}
\begin{center}
\begin{tabular}{ccccccc}
\hline
 Symmetry     & $A_1$ & $A_2$ & $B_1$ & $B_2$ & $C_1$ & $C_2$ \\
\hline
\hline
$U(N)_{1,k}$& $0$ & $0$ & $\bar N$ & $0$ & $0$ & $N$ \\
$U(N)_{2,k}$& $0$ & $0$ & $0$ & $N$ & $\bar N$ & $0$ \\
$U(N)_{3,-k}$& $\bar N$ & $\bar N$ &$N$ & $0$ & $N$ & $0$ \\
$U(N)_{4,-k}$& $N$ & $N$ &$0$ & $\bar N$ & $0$ & $\bar N$ \\ 
\hline
$U(1)_1$ & $\frac{1}{2}$ & $-\frac{1}{2}$ & $0$ & $0$ & $0$ & $0$ \\
$U(1)_2$ & $0$ & $0$ & $\frac{1}{2}$ & $-\frac{1}{2}$ & $0$ & $0$ \\
$U(1)_3$ & $0$ & $0$ & $0$ & $0$ & $\frac{1}{2}$ & $-\frac{1}{2}$ \\
$U(1)_R$ & $1-2h$ & $1-2h$ & $h$ & $h$ & $h$ & $h$ \\
\hline
\end{tabular}
\end{center}
\end{table}
The superpotential of the theory is
\begin{equation}
W=\tr(\epsilon^{ij}C_2B_1A_iB_2C_1A_j). 
\end{equation}
It turns out that the moduli space of the theory is $Q^{1,1,1}/{\bf Z}_k$,
and this theory is expected to be dual to M-theory on AdS$_4\times Q^{1,1,1}/{\bf Z}_k$. 
Here ${\bf Z}_k$ is included in the diagonal $SU(2)$ of
$SU(2)_2\times SU(2)_3$, 
so ${\bf Z}_k$ breaks $SU(2)_2\times SU(2)_3$ to $U(1)_F\times U(1)_B$. 
The manifest global symmetry in the Lagrangian is
\begin{equation}
SU(2)_1\times U(1)_F\times U(1)_B\times U(1)_R.
\label{symq111}
\end{equation}
The charge assignment of $U(1)_B (U(1)_F)$ is
given by the summation (difference) of those of $U(1)_2$ and $U(1)_3$.
Using the symmetry and the fact that 
the superpotential has the R-charge $2$, 
we can determine R-charges of the chiral fields with one parameter $h$.
If AdS$_4$/CFT$_3$ duality is the case in this model, 
then the gauge theory must have the global symmetry \eqref{isom} 
rather than \eqref{symq111} when $k=1$ in the large $N$ limit. 
In the following, 
we give  evidence of this global symmetry enhancement by using a superconformal index.

Since we fix the charge assignments of the fields, 
we can compute an index of this theory by applying the formula.
We compute the index of this model numerically up to $x^2$ order when $k=1$.
The neutral part of the index is 
\begin{equation}
I^{(0)}(x,z_i)=1+
\chi_\frac{1}{2}(z_1)
\left(\frac{z_2^{1/2}}{z_3^{1/2}}+\frac{z_3^{1/2}}{z_2^{1/2}}\right)
x+
\left[
\chi_1(z_1)\left(2\frac{z_2}{z_3}+1+2\frac{z_3}{z_2}\right)
-3
\right]
x^2+\cdots,
\label{q111i0}
\end{equation}
where $\chi_s(z)$ is the $SU(2)$ character with spin $s$
\be
\chi_s(z)=z^s+z^{s-1}+\cdots+ z^{-s}.
\ee
The charged parts are 
\bea
\!\!\!\!\!\!\!\!I^{(+)}&=&1+\chi_\frac{1}{2}(z_1)z_2^{-1/2}z_3^{-1/2}x
+\left[(\chi_1(z_1)-1)(z_2^{-1}+z_3^{-1})
+2\chi_1(z_1)z_2^{-1}z_3^{-1}\right]x^2
+\cdots.
\label{q111ip}\\
\!\!\!\!\!\!\!\!I^{(-)}&=&1+\chi_\frac{1}{2}(z_1)z_2^{1/2}z_3^{1/2}x
+\left[(\chi_1(z_1)-1)(z_2+z_3)
+2\chi_1(z_1)z_2z_3\right]x^2
+\cdots.
\label{q111im}
\eea
To obtain the positive part of index, 
we sum up the following contributions including $I^{(+)}_{(0,0,0,0)}(x,z_i)=1$. 
\begin{eqnarray}
I^{(+)}_{(1,1,1,1)}(x,z_i)
&=&\chi_\frac{1}{2}(z_1)z_2^{-1/2}z_3^{-1/2}x
+(\chi_1(z_1)-1)(z_2^{-1}+z_3^{-1})x^2
+\cdots,
\nonumber\\
I^{(+)}_{(2,2,2,2)}(x,z_i)
&=&\chi_1(z_1)z_2^{-1}z_3^{-1}x^2
+\cdots,
\nonumber\\
I^{(+)}_{((1,1),(1,1),(1,1),(1,1))}(x,z_i)
&=&\chi_1(z_1)z_2^{-1}z_3^{-1}x^2+\cdots.
\end{eqnarray}

We have several remarks from these results.
First,  we observe that 
the index does not depend on $h$.
Second, 
we can easily see that the index has the following properties. 
\begin{eqnarray}
I^{(*)}(x,z_1,z_2,z_3)&=&I^{(*)}(x,z_1^{-1},z_2,z_3),
\label{q111isym1}\\
I^{(*)}(x,z_1,z_2,z_3)&=&I^{(*)}(x,z_1,z_3,z_2),
\label{q111isym2}\\
I^{(-)}(x,z_1,z_2,z_3)&=&I^{(+)}(x,z_1,z_2^{-1},z_3^{-1}).
\label{q111isym3}
\end{eqnarray}
And each relation comes from symmetry underlying the theory. 
(\ref{q111isym1}) comes from $SU(2)_1$ symmetry. 
(\ref{q111isym2}) is due to $\bZ_2$ symmetry, 
which exchanges $B_i$ and $C_i$. 
(\ref{q111isym3}) originates in the charge conjugation symmetry, 
which exchanges $B_1, C_1$ and $B_2, C_2$, respectively.

To see the flavor symmetry enhancement to $SU(2)^3$,
we compute the total index by using the factorization property \eqref{factor}. 
To do this, 
we simply multiply (\ref{q111i0}), (\ref{q111ip}), and (\ref{q111im}). 
\begin{equation}
I(x,z_i)
=1
+\chi_{\frac{1}{2}}(z_1)\chi_{\frac{1}{2}}(z_2)\chi_{\frac{1}{2}}(z_3)x
+(2\chi_1(z_1)\chi_1(z_2)\chi_1(z_3)-2)x^2
+\cdots. 
\label{indexq111}
\end{equation}
This is invariant under the Weyl reflections
$z_i\rightarrow z_i^{-1}$ and under permutations among $z_i$, 
which implies that the theory has $SU(2)^3$ symmetry.
This is precisely what we expect from the isometry of $Q^{1,1,1}$.

The comparison with the index of dual M-theory was carried out in \cite{Cheon:2011th}.
The gravity index can be obtained by summing the BPS spectrum 
of M-theory on AdS$_4\times Q^{1,1,1}$, 
which was investigated in \cite{Merlatti:2000ed}. 
It turns out from the spectrum that up to the $x^2$ order
the single particle index receives the contribution only from one hypermultiplet and short vector multiplets. 
The result is 
\begin{eqnarray}
&&
I^{\rm sp}(x,z_i)=
\chi_\frac{1}{2}(z_1)
\chi_\frac{1}{2}(z_2)
\chi_\frac{1}{2}(z_3)x
\nn
&&\qquad+
\left(
\chi_1(z_1)
\chi_1(z_2)
\chi_1(z_3)
-\chi_1(z_1)
-\chi_1(z_2)
-\chi_1(z_3)
-2
\right)x^2+\cdots.
\label{q111sp}
\end{eqnarray}
Note that the positive terms are coming from the hypermultiplet and 
the negative terms are from (massless) vector multiplets. 
One can easily see that 
the multi-particle index for M-theory on AdS$_4\times Q^{1,1,1}$, 
which can be obtained as plethystic exponential of (\ref{q111sp}), 
agree with \eqref{indexq111} up to the order of $x^2$. 

Consult \cite{Imamura:2011uj,Cheon:2011th} for further investigation
such as applications to other ${\cal N}=2$ Chern-Simons matter theories 
and their gravity duals.

\section{Summary and discussion}

We have seen aspects of three dimensional superconformal indices. 
They can be computed exactly 
by using localization method including the contribution of monopole operators, 
which play a key role to elucidate the rich structure of 
three-dimensional superconformal field theories and their M-theory duals. 
We showed that superconformal indices can capture the rich structure 
and non-perturbative aspects such as mirror duality, AdS$_4$/CFT$_3$ duality.

In Section \ref{ads/cft}, 
using a superconformal index 
we showed  evidence of a global symmetry enhancement 
of a gauge theory dual to M-theory on AdS$_4\times Q^{1,1,1}$
in $k=1$ in the large $N$ limit. 
It is natural to ask whether this phenomenon occur in a finite $N$. 
We strongly suspect this is not the case at least in this model.  
This is simply due to the fact that 
in a finite $N$ case 
a non-diagonal monopole operator contributes to the index non-trivially 
and thus the enhancement doesn't happen. 
To discuss this concretely, 
let us compute the index of the gauge theory in abelian case.
This can be done simply 
by applying the formulas \eqref{formula}, \eqref{fvecdef} and \eqref{fchdef}. 
It turns out that 
several non-diagonal monopole operators contribute to the index non-trivially like 
\begin{eqnarray}
I_{(1,-1,0,0)}(x,z_i)
&=&x^4-\biggl(\frac{1}{z_1^2}+z_1^2\biggr)x^6+\cdots,\\
I_{(0,0,-1,1)}(x,z_i)
&=&x^{2+2h}-4 x^{6+ 2h}
+\chi_\half(z_1)\biggl({z_2^2\over z_3^2}+{z_3^2 \over z_2^2}\biggr)x^{7+2h}
+\cdots. 
\end{eqnarray}
Contributions coming from non-diagonal monopole operators like these
break the invariance under the Weyl reflections and permutations of $z_i$.%
\footnote{
The contribution of the index only from diagonal monopole operators up to the order of $x^3$
is 
\bea
\!\!\!\!\!\!\!\!\!\!\!\!\!\!\!\!\!\!\!\!\!\!\!\!&&I_{diagonal}(x,z_i)
=1+\chi_{\frac{1}{2}}(z_1)\chi_{\frac{1}{2}}(z_2)\chi_{\frac{1}{2}}(z_3)x 
+(\chi_1(z_1)\chi_1(z_2)\chi_1(z_3)-\chi_1(z_1)-\chi_1(z_2)-\chi_1(z_3)-3)x^2 \nn
\!\!\!\!\!\!\!\!\!\!\!\!\!\!\!\!\!\!\!\!\!\!\!\!&&\qquad+\chi_\half(z_1)\chi_\half(z_2)\chi_\half(z_3)
(\chi_1(z_1)\chi_1(z_2)\chi_1(z_3)-\chi_1(z_1)\chi_1(z_2)-\chi_1(z_2)\chi_1(z_3)-\chi_1(z_1)\chi_1(z_3)+2)x^3 \nn
\!\!\!\!\!\!\!\!\!\!\!\!\!\!\!\!\!\!\!\!\!\!\!\!&&\qquad +\cO(x^4), 
\label{indq111finite}
\eea
which is invariant under the Weyl reflections and permutation of $z_i$.
}

Therefore, we naturally expect that 
these contributions from non-diagonal monopole operators decouple 
after the large $N$ limit. 
This expectation is also natural from the dual geometry perspective. 
We proposed in \cite{Imamura:2008ji} 
that 
a non-diagonal monopole operator corresponds to 
a M2-brane wrapped on a two-cycle in an internal manifold. 
We gave evidence for this on $\cN=4$ Chern-Simons theories 
by using supconformal indices \cite{Imamura:2009hc,Imamura:2010sa}. 
In case of $\cN=4$ Chern-Simons theories, 
such two cycles are vanishing ones coming from orbifold singularities, 
so wrapped M2-branes on them are BPS and contribute to the index
of order $1$. 
In a general $\cN=2$ Chern-Simons theories, however, 
such two cycles have a finite volume.
Therefore, even if wrapped M2-branes on them become BPS and 
contribute to the index,
the contribution is of order $x^{\sqrt{N}}$, 
so decouple after the large $N$ limit.

It is worth mentioning that 
this consideration is not inconsistent with 
the earlier study that 
the supersymmetry is enhanced from $\cN=6$ to $\cN=8$ in ABJM theory with $k=1, 2$
\cite{Gustavsson:2009pm,Kwon:2009ar,Bashkirov:2010kz,Samtleben:2010eu}. 
This is simply because
ABJM theory does not have non-diagonal monopole operators \cite{Imamura:2008ji}. 
Indeed, for example, 
we can compute the index of abelian ABJM theory with $k=1$ by using the formula.
Here we use the normalization in \cite{Kim:2009wb}. 
We can evaluate it analytically and the result is given by the plethystic exponential of the following single-particle index 
\bea
\mfi_{ABJM} &=& 
(v^\half \chi_{1/2}(z_A) y+ v^{-\half} \chi_{1/2}(z_B) y^{-1})
{x^{\half} \over {1-x^2}} \nn
&&- (v^\half \chi_{1/2}(z_A)  y^{-1} + v^{-\half} \chi_{1/2}(z_B)y)
{x^{\frac{3}{2}} \over 1-x^2}.
\label{abjmsp}
\eea
Here $z_A$ and $z_B$ are chemical potentials for $SU(2)_A$
and $SU(2)_B$ symmetries rotating the complex scalars $A_1, A_2$ and $B_1, B_2$, respectively. 
$v$ is a chemical potential for $U(1)_b$ symmetry and 
$y$ is that for the diagonal monopole charge.
By setting $v=1$, which means making the baryonic charge of the fields trivial, 
this single-particle index \eqref{abjmsp}
reduces to that of four complex chiral multiplets with a suitable charge assignment. 
This is consistent with the expectation that abelian ABJM theory is
an effective field theory of one membrane, which, in particular, has $SO(8)$ R-symmetry.

On the other hand,  
a discrepancy was found between 
an index for several $\cN=2$ gauge theories calculated by using the large $N$ formula 
and that for their gravity duals calculated from the BPS spectrum obtained by Kaluza-Klein analysis at a higher order \cite{Cheon:2011th}. 
It will be worthwhile to investigate whether there is a discrepancy in other $\cN=2$ models, 
and what is the problem if there is.

We hope we will come back to these issues in the near future.

\vspace*{1ex}
\section*{Acknowledgments}
The auther would like to thank Yosuke Imamura and Daisuke Yokoyama
for early collaboration. 
He is also grateful to Yosuke Imamura and Seok Kim for nice comments. 
He was supported by National Research Foundation of Korea (NRF) grant No. 2010-0007512,
and No. 2009-0076297.

\bibliographystyle{utphys}
\bibliography{ref-bib}

\end{document}